\documentstyle[sprocl]{article}

\def\Journal#1#2#3#4{{#1} {\bf #2}, #3 (#4)}

\def\NPB{{\em Nucl. Phys.} B}
\def\PLB{{\em Phys. Lett.}  B}
\def\PRL{\em Phys. Rev. Lett.}
\def\PRD{{\em Phys. Rev.} D}

\def\be{\begin{equation}}
\def\ee{\end{equation}}
\def\bea{\begin{eqnarray}}
\def\eea{\end{eqnarray}}
\font\bigbf=cmssbx10 scaled\magstep2
\def\etal{{\it et al,\/}\ }

\begin{document}

\rightline{UFIFT--HEP--98/11}
\rightline{hep-ph/9807341}
\rightline{July 1998}
\title{The $Z_2\times Z_2$ Orbifold and the SUSY Flavor Problem
\footnote{Invited talk presented at PASCOS 98, Boston MA, March 22--29 1998.}
}
\author{Alon E. Faraggi}

\address{Department of physics, University of Florida, Gainesville, FL 32611,
USA\\E-mail: faraggi@phys.ufl.edu} 


\maketitle\abstracts{ 
The characteristic property of the $Z_2\times Z_2$ orbifold
is its cyclic permutation symmetry. It is argued that this
property may be instrumental in explaining simultaneously
the fermion mass hierarchy and the squark mass degeneracy.
Detailed studies in free fermionic models
that preserve the cyclic permutation symmetry of anomalous
$U(1)$ charges of the three generations are discussed. 
}
\section{The SUSY Flavor Problem }
The flavor problem in supersymmetric extensions of the
Standard Model is especially interesting. On the
one hand flavor dependent symmetries are motivated
by the need to explain the hierarchical fermion mass spectrum.
On the other hand, the absence of Flavor Changing Neutral
Currents at an observable rate suggests the need for
flavor independent symmetries, which force squark mass
degeneracy. Thus, the natural question arises how can
these, seemingly orthogonal, symmetries coexist.
In supersymmetric, two doublet Higgs models,
the first problem requires that the Yukawa couplings
which couple the fermions to the Higgs fields exhibit
an hierarchical pattern. 
The second problem requires that the soft SUSY breaking
parameters are either highly degenerates, or are 
aligned with the ordinary quark Yukawa matrices.
In the context of supersymmetric field theories,
the various parameters are fixed to agree with the data,
and acceptable solutions can of course be found. 
The problem, however, becomes more interesting
when one tries to unify the gauge interactions with
gravity. Superstring models provide useful laboratories
to study such theories. In this case the flavor symmetries,
in a given model, are imposed and cannot be chosen
arbitrarily. While the study of string models is
still in its infancy, it is in general indeed expected 
that superstring models give rise to non--universal 
soft SUSY breaking parameters \cite{il}. Therefore, string models
that can on the one hand give rise to flavor dependent
symmetries that are needed to explain the fermion
flavor mass pattern while, on the other hand producing
a mechanism to produce nearly flavor degenerate
soft SUSY breaking parameters, may be of special
interest. Furthermore, recent progress in understanding
nonperturbative aspects of string compactifications
suggests that the desired symmetries should also
remain in the nonperturbative extension of 
potentially realistic superstring models,
which are constructed at weak coupling. 
\section{Realistic Free fermion Models} 
In this talk I discuss a class of string models
that can indeed resolve the fermion mass
puzzle on the one hand, while providing
a robust symmetry structure that may,
at least partially, alleviate the 
sfermion flavor problem. We have to caution
however that a completely mutually satisfactory
solution to both problems, in the context
of string solutions, can only be obtained
once the full fermion spectrum is
computed quantitatively and similarly
for the sfermions soft masses. However,
short of this ambitious, and still distant,
goal, we can still observe qualitative properties
of specific string compactifications and suggest scenarios
how both problems can be resolved in such an eventual
calculation.

There are several possible ways to try to construct realistic
superstring models. One possibility is to construct models
with a GUT gauge group, like $SU(5)$
$SO(10)$ or $E_6$, which is broken to the Standard Model
gauge group at an intermediate energy scale \cite{stringguts}.
Another possibility
is to construct superstring models with semi--simple GUTs,
like $SU(3)^3$ \cite{suthree}, 
$SU(5)\times U(1)$ \cite{revamp} or
$SO(6)\times SO(4)$ \cite{patisalamstrings}.
Finally, we can construct superstring models in which
the non--Abelian factors of the Standard Model are obtained
directly at the string level \cite{zthree,eu,slm}.
A realistic model of unification
must satisfy a large number of restrictive phenomenological
constraints, a few of which are listed below.

\centerline{{$\underline{{\hbox{~~~~~~~~~~~~~~~~~~~~~~~~~~~~~~~~~~~~~}}}$}}

{}~~~~~$1.$ Gauge group ~$\longrightarrow$~ $SU(3)\times SU(2)\times U(1)_Y$

{}~~~~~$2.$ Contains three generations

{}~~~~~$3.$ Proton stable ~~~~~~~~($\tau_{\rm P}>10^{30+}$ years)

{}~~~~~$4.$ Agreement with $\underline{\sin^2\theta_W}$ and
$\underline{\alpha_s}$ at $M_Z$ 

{}~~~~~$5.$ N=1 supersymmetry~~~~~~~~(or N=0)

{}~~~~~$6.$ Contains Higgs doublets $\oplus$ potentially realistic
Yukawa couplings

{}~~~~~$7.$ Light left--handed neutrinos

{}~~~~~~~~~~~~~$~8.$ $SU(2)\times U(1)$ breaking

{}~~~~~~~~~~~~~$~9.$ SUSY breaking

{}~~~~~~~~~~~~~$10.$ No flavor changing neutral currents

{}~~~~~~~~~~~~~$11.$ No strong CP violation

{}~~~~~~~~~~~~~$12.$ Exist family mixing and weak CP violation

{}~~~~~$13.$ +~~ {\bf ...}

{}~~~~~$14.$ +~~~~~~~~~~~~~~~~{\bigbf{GRAVITY}}

\centerline{{$\underline{{\hbox{~~~~~~~~~~~~~~~~~~~~~~~~~~~~~~~~~~~~~}}}$}}

\smallskip

The last criteria provides a strong motivation for the study
of superstring models.
It is often stated that superstring theory contains an enormous
number of consistent models and we have no mechanism to select
among these vacua. This is especially significant in view
of the fact that even the deeper understanding of nonperturbative
aspects of string theory, gained in recent years,
does not seem to lessen the problem.
However, this notion is very misleading. Nearly all of these
consistent vacua do not satisfy even the first two necessary requirements.
Moreover, many of the models, that have been labeled as ``realistic'',
satisfy {\it only} the first two criteria, plus possibly
existence of possible light Higgs doublets and couplings
between the fermions and Higgs doublets that can be identified as
Yukawa couplings. However, as we go down the list
nearly all string models that have been constructed to date are
excluded. Take, for example, the fourth requirement in the 
list above, and consider the three generation standard--like
orbifold models. Such three generation models were 
constructed in the $Z_3$, $Z_7$ and $Z_2\times Z_2$
orbifold. In the first two cases one finds a weak--hypercharge
embedding with the normalization $k_Y>5/3$, whereas
the last one produced three generation models with the
standard $SO(10)$ embedding, {\it i.e.} with $k_Y=5/3$.
It is precisely for this reason that the superstring
models, which are based on the $Z_2\times Z_2$ orbifold,
can be in agreement with the measured values of $\alpha_s(M_Z)$
and $\sin^2\theta_W(M_Z)$. It is important to emphasize
that this does not mean that one of string models that have
constructed to date is necessarily the correct string vacuum.
Rather, it shows that the $Z_2\times Z_2$ orbifold 
compactification contains the desired ingredients that
can explain many of the phenomenological requirements
in the above list. Thus motivating the hypothesis that
the true string vacua is a $Z_2\times Z_2$ orbifold, in the
vicinity of the free fermionic models. 

The superstring models, based on a $Z_2\times Z_2$ orbifold
compactification have been studied in the framework
of the free fermionic construction \cite{FFF}, in which the moduli
of the compactified dimensions are frozen at the value,
where they can be realized as free fermions propagating
on the world--sheet. The models are specified in terms
of a set of boundary condition basis vectors for
all world--sheet fermions, which are constrained by the 
string consistency requirements. The physical
spectrum is obtained by applying the generalized GSO
projections. The low energy effective field theory
is obtained by S--matrix elements between external states.
The Yukawa couplings and higher order nonrenormalizable
terms in the superpotential are obtained by calculating
correlators between vertex operators. For a correlator
to be non vanishing all the symmetries of the model
must be conserved. The boundary condition basis vectors
completely define the spectrum and the symmetries of
a given string model, and therefore its phenomenological
properties. 

The first five basis vectors $\{{\bf 1},S,b_1,b_2,b_3\}$,
in the models of interest here,
consist of the so called NAHE
set, which yields after the generalized GSO projections
an $N=1$ supersymmetric $SO(10)\times SO(6)^3\times E_8$
gauge group. The three sectors $b_1$, $b_2$ and $b_3$
correspond to the three twisted sectors of the $Z_2\times Z_2$
orbifold. The correspondence is explicitly demonstrated by
adding to the NAHE set the basis vector $X$ with periodic
boundary conditions for the right--moving world--sheet fermions
$\{{\bar\psi}^{1,\cdots,5},{\bar\eta}^1,{\bar\eta}^2,
{\bar\eta}^3\}$ \cite{ztwo,efn}.
The gauge group in this case is $E_6\times U(1)^2\times SO(4)^3\times E_8$,
with 24 generations in the 27 of $E_6$. The same model
is constructed in the bosonic language by specifying
the background metric and antisymmetric tensor field
and then moding out by the $Z_2\times Z_2$ discrete symmetry.
In the realistic models the
sign of the GSO phase $c(X;\xi={\bf1}+b_1+b_2+b_3)$
is flipped, which breaks $E_6\rightarrow SO(10)\times U(1)$,
and $E_8\rightarrow SO(16)$.
The $27$ representations of $E_6$ now become 24 representations
in the chiral 16 of $SO(10)$, plus 24 representations in
the vectorial 16 of $SO(16)$. The $U(1)$ current
that was previously embedded in $E_6$ now becomes
an anomalous, flavor independent, $U(1)$ symmetry.
The two orthogonal $U(1)$ combinations are, flavor
dependent, anomaly free $U(1)$ currents.

There are two features that are important to
note from the discussion above. The first is the appearance
of anomalous $U(1)$ symmetry, as a result of the breaking
of $E_6\rightarrow SO(10)\times U(1)^2$, which is family
universal. The second is the cyclic permutation symmetry
between the three sectors $b_1$, $b_2$ and $b_3$
with respect to their left-- and right--moving
quantum numbers under the world--sheet currents.
This cyclic permutation symmetry is the characteristic
property of the $Z_2\times Z_2$ orbifold compactification,
with the standard embedding of the gauge connection.

The massless spectrum of the $E_6$ model contains also
three pairs of $27\oplus{\overline{27}}$ from the untwisted
sector. In the $SO(10)\times U(1)$ model these states are broken
to three pairs of the vectorial $10$ representation of $SO(10)$.
These states also respect the cyclic permutation symmetry of the
$Z_2\times Z_2$ orbifold, and give rise to the Yukawa couplings
$16_i16_i10_i$, where $i$ denotes the $i^{th}$ twisted sector. 

The three generation free fermionic models are constructed
by adding three additional boundary condition basis vectors 
beyond the NAHE set \cite{slm}. These break the $SO(10)$ and $E_8$ gauge
groups to one of their subgroups. The flavor $SO(6)$ symmetries
are broken to a product of $U(1)$ factors. The twisted sectors
give three generations, one from each $b_1$, $b_2$ and $b_3$.
The vectorial untwisted representations produce three pairs
of Higgs representations. The additional vectors, beyond
the NAHE set, give rise to additional massless spectrum
which varies between models. These representations
are in general vector--like. In all models that have been studied
in detail there is one vector combination which 
produces additional light Higgs representations.
These additional Higgs representations play
an important role in the analysis of fermion mass
matrices. The other states from the sectors beyond the
NAHE set are mostly exotic vector--like states with
fractional electric charge or fractional $U(1)_{Z^\prime}$.
For appropriate choices of flat directions these states
can decouple from the massless spectrum and are not of
interest here. 

In general, the additional boundary conditions basis vectors,
beyond the NAHE set break the cyclic permutation symmetry. 
What is remarkable is that in some models the cyclic permutation
symmetry of the three generations with respect to their
$U(1)$ quantum numbers\footnote{before the anomalous and
anomaly free combinations are taken.} is preserved.
As a result in these models the
anomalous $U(1)$ is {\it family universal}.
Interestingly, the boundary condition basis vectors
that preserve this cyclic permutation symmetry have
some specific properties \cite{CF,univ}.
\section{Fermion Mass Hierarchy}
We can now start to see how, with the above general structure and
in the presence of a
family universal anomalous $U(1)$ both the fermion and sfermion
flavor problem can be resolved simultaneously.
The details of the analysis has been presented elsewhere \cite{CKM,FP2},
so only the qualitative structure is discussed here.
The fermion mass terms are of the form 
$cgf_if_jh\phi^{^{N-3}}$ or
$cgf_if_j{\bar h}\phi^{^{N-3}}$, where $c$ is a 
calculable coefficient, $g$ is the gauge coupling at the unification 
scale,  $f_i$, $f_j$ are the fermions from
the sectors $b_1$, $b_2$ and $b_3$, $h$ and ${\bar h}$ are the light 
Higgs doublets, and $\phi^{N-3}$ is a string of Standard Model singlets
that get a VEV and produce a suppression factor
${({{\langle\phi\rangle}/{M}})^{^{N-3}}}$ relative to the cubic
level terms. 
Each generation from a sector $b_j$ is charged with respect
to the left--moving, $U(1)_{L_j}$ and  $U(1)_{L_{j+3}}$,
and right--moving $U(1)_{R_j}$ and  $U(1)_{R_{j+3}}$, symmetries.
Each Higgs doublet pair from the NS sector is charged under
$U(1)_{R_j}$. Consequently at the cubic level only the couplings
$\{u_jQ_j+N_jL_j\}{\bar h}_j$ and $\{d_jQ_j+e_jL_j\}h_j$
are allowed. Note that each generation couples to a different
Higgs pair, and that at this level the cyclic permutation symmetry
is retained. As the anomalous $U(1)$ Fayet--Iliopoulos term
breaks supersymmetry
near the Planck scale, we must assign VEVs to some Standard
Model singlets, along flat $F$ and $D$ directions. In this process
some of the nonrenormalizable terms become effective renormalizable
operators \cite{flatanalysis}.
At the same time some of the Higgs doublet representations
receive large mass. For specific solutions only two Higgs doublets
remain massless down to the electroweak scale. We should warn
again that the full solution to the Higgs mass spectrum
can only be obtained once we have a full solutions
to the fermion mass spectrum. That is VEVs that enter the
fermion mass terms also affect the Higgs mass matrix.
For our purpose here we can make the assumption that the
light Higgs spectrum consist of two Higgs doublets and
examine the qualitative pattern of the fermion mass
matrices that arises. Analysis of the nonrenormalizable terms
up to order $N=8$ reveals the following structure
$${M_U\sim\left(\matrix{\epsilon,a,b\cr
                    {\tilde a},A,c \cr
                    {\tilde b},{\tilde c},\lambda_t\cr}\right);{\hskip .2cm}
M_D\sim\left(\matrix{\epsilon,d,e\cr
                    {\tilde d},B,f \cr
                    {\tilde e},{\tilde f},C\cr}\right);{\hskip .2cm}
M_E\sim\left(\matrix{\epsilon,g,h\cr
                    {\tilde g},D,i \cr
                    {\tilde h},{\tilde i},E\cr}\right)},$$
where $\epsilon\sim({{\Lambda_{Z^\prime}}/{M}})^2$.
The diagonal terms in capital letters represent leading 
terms that are suppressed by singlet VEVs, and $\lambda_t=O(1)$.
The mixing terms are generated by hidden sector states from the
sectors $b_j+2\gamma$ and are represented by small letters. They 
are proportional to $({{\langle{TT}\rangle}/{M}^2})$.
The states from the sector $b_3$ are identified with the
lightest generation.

The important new feature which arises from this analysis is that
the Higgs spectrum at the string scale is more complicated
than a simple two Higgs doublet model. Then although there
exist a permutation symmetry at the level of the NAHE set
this permutation symmetry interchanges between the twisted
sectors, as well as between the untwisted Higgs doublets.
The permutation symmetry is partially broken by the additional
boundary condition basis vectors, beyond the NAHE set, and is broken
further by the choices of flat directions. Consequently,
only two of the Higgs representations, say ${\bar h}_1$ and ${h_{45}}$,
remain light. Only the top couples to $\bar h_1$ at the cubic level
and the mass terms for the lighter quarks and leptons arise
from, successively suppressed, higher order terms.
\section{Sfermion Mass Degeneracy}
Although the permutation symmetry is broken in the fermion mass
sector, the charges with respect to the anomalous $U(1)$ are
family universal. This is the case, for example, in the model
of ref. \cite{eu}. This offers an intriguing possibility.
If the dominant source of supersymmetry breaking is the
anomalous $U(1)$ then the resulting soft squark masses
are family universal. Supersymmetry breaking will occur,
at hierarchically small scale if there is a mass term,
$m\Phi{\bar\Phi}$,
for some Standard Model singlet, which is charged under the
anomalous $U(1)$. The effective potential then takes the form
\be
V={g^2\over2}\sum_\alpha D_\alpha^2+m^2(\vert\Phi\vert^2
+\vert{\bar\Phi}\vert^2)
\ee
where $D_\alpha$ are the various $U(1)$ $D$--terms,
and we assumed a common coupling $g$ at the unification scale,
to simplify the analysis. Extremizing the potential
it is found that SUSY is broken. Furthermore, for a specific 
solution of the $F$ and $D$ flatness constraints
it is found that the mass term $m$ is hierarchically suppressed
and that in the minimum the $D$--terms of the family universal
$U(1)$'s are nonzero, whereas those of the family dependent
$U(1)$ vanish. This solution therefore provides an example
how the squark mass degeneracy may arise, provided that the
dominant component that breaks supersymmetry is the anomalous
$U(1)$ D--term. 
Furthermore, the mass term $m$, which breaks supersymmetry,
can be hierarchically small relative to the Planck scale.
This is because such a term must arise from
nonrenormalizable terms that contain hidden sector
condensates. The condensation scale in the hidden sector
is determined by its gauge and matter content.
For example, in the model of ref \cite{eu} we found a cubic
level flat $F-D$ solution, with the mass term $m$ induced
at order $N=8$, by matter condensates of the hidden $SU(5)$
gauge group \cite{FP2}. 

The general fermion mass pattern that we found,
as well as the flavor universality of the anomalous $U(1)$ are
an intrinsic reflection of the underlying $Z_2\times Z_2$ orbifold 
structure. Therefore, we see that the $Z_2\times Z_2$
orbifold contains the intrinsic structure that is needed
in order to understand both the fermion mass spectrum
as well as the sfermion mass degeneracy. We should remark,
however, that we used different flat solutions to analyze
each of the scenarios separately. Whether or not it is
possible to find a solution that produces the qualitative
fermion mass pattern as well as the sfermion mass degeneracy
{\it simultaneously}, still remains to be seen. 

SUSY breaking by a family universal anomalous $U(1)$ can
produce the desired squark degeneracy. However,
the gaugino masses can (at best) only arise at the one
loop level in string perturbation theory. In this case
the gluino masses are approximately give by
$
m({\lambda_a})\approx \lambda^\prime
\langle F({\bar\Phi})\langle\Phi\rangle/
M_{Pl}^2\approx \lambda^\prime\epsilon m
\label{gauginomasses}
$,
where $\lambda^\prime\le{\cal O}(1)$ and $\epsilon\approx1/50$.
This induces the hierarchy\cite{FP2}
\be
[m^2(\tilde q_i)\approx Q_A^i(\sqrt{3}m^2/2)]~>~
[\Delta m_{\tilde q_i}^2\approx
\lambda\epsilon(m^2/2)]~>~ [m_{\lambda_a}^2\approx\lambda^\prime\epsilon^2m^2]
\label{sqm2}
\ee
Because of this hierarchy, it is clear that if SUSY breaking
proceeds entirely through anomalous $U(1)$, the gluinos typically
would be rather light. From (\ref{gauginomasses}),
one obtains: $m_{\tilde g}\approx 2\lambda^\prime\epsilon m_{\tilde q}
\approx \lambda^\prime(20-60){\rm GeV}$, for $m_{\tilde q}\approx
(1-3){\rm TeV}$; this may be too light,
compared to the observed limit on $m_{\tilde g}$ of 130GeV,
unless $\lambda^\prime\ge2$
and $m_{\tilde q}\ge3{\rm TeV}$. To make matters worse, for string
solutions, as considered here, $\lambda^\prime$ vanishes at tree
level and can only arise through quantum loops; thus it is expected
to be small. This suggests that SUSY
breaking through anomalous $U(1)$, quite plausibly, is accompanied
by an additional source which provides the dominant contribution to
gluino masses $(\sim(1-~{\rm few})(100{\rm GeV}))$, while preserving
the squark--degeneracy, obtained through $U(1)_A$.
The interesting possibility is that of SUSY breaking through
a combined dilaton--anomalous $U(1)$ scenario.
Each of this scenarios separately encounters some difficulties.
That of the anomalous $U(1)$ has the problem of small gaugino
masses, whereas the dilaton dominated scenario has the problem
with color and electric charge breaking.
Such a combined analysis may also be instrumental
in trying to resolve the dilaton stabilization problem \cite{dsp}. 

We have seen that the $Z_2\times Z_2$ orbifold has
an intriguing structure, which is very appealing
from the point of view of the fermion and sfermion masses.
Several other properties of this class of superstring
compactification are worth noting. 
The first is the fact that the chiral generations
from the twisted sectors $b_1$, $b_2$ and $b_3$
carry positive charges under the anomalous $U(1)$.
This is an important property, as otherwise SUSY breaking
through the anomalous $U(1)$ would lead to color and electric
charge breaking. This positivity of the squarks and sleptons
charges under $U(1)_A$ is again a reflection of the $Z_2\times Z_2$
orbifold structure and the partial embedding of the anomalous $U(1)$
in $E_6\rightarrow SO(10)\times U(1)_A$.
The string models contain also color, and electrically, charged,
vector--like states, that may have negative $U(1)_A$
charge, However, in all the cases that we examined these
states receive large intermediate mass which over--compensates
for the $U(1)_A$ $D$--term contribution. 
It is also important to note that the
universality structure of the $Z_2\times Z_2$ orbifold
is also reflected in other sectors
of the theory. In particular for the untwisted moduli. 
Thus, even if SUSY--breaking is dominated by the untwisted
moduli sector squark degeneracy is still expected.
\section{Conclusions}
String theory provides a window to study the unification
of the gauge interactions with gravity. To bring this
exploration to contact with experimental physics
necessitates the construction of phenomenologically realistic
models. The free fermionic models provide examples of such
phenomenologically appealing models. Underlying these
models there is a special $Z_2\times Z_2$ orbifold.
In this talk I discussed how the characteristic property
of the $Z_2\times Z_2$ orbifold, namely its cyclic permutation
symmetry, may play a pivotal role in simultaneously understanding
the fermion mass hierarchy and squark mass degeneracy.

\section*{Acknowledgments}
It is a pleasure to thank Gerald Cleaver and Jogesh Pati
for collaboration on the work reported in this talk.
Work supported in part by DOE grant No. DE--FG--0586ER40272.


\section*{References}
\begin{thebibliography}{99}
\bibitem{il}L.E. Ibanez and D. Lust, \Journal{\NPB}{382}{305}{1992}.
\bibitem{stringguts}D.C. Lewellen, \Journal{\NPB}{337}{61}{1990};
        Z. Kakushadze and S.H.H. Tye, \Journal{\PRL}{77}{2612}{1996};
                                        \Journal{\PRD}{55}{7896}{1997}.
\bibitem{suthree}P. Candelas \etal
                                        \Journal{\NPB}{258}{46}{1985};
                B. Greene \etal
                                        \Journal{\NPB}{292}{606}{1987};
                R. Arnowitt and  P. Nath \Journal{\PRL}{62}{2225}{1989};
\bibitem{revamp}I. Antoniadis \etal \Journal{\PLB}{231}{65}{1989};
		J. Lopez \etal \Journal{\NPB}{399}{654}{1993}.
\bibitem{patisalamstrings} I. Antoniadis \etal \Journal{\PLB}{245}{161}{1990};
                G.K. Leontaris \Journal{\PLB}{372}{212}{1996}.
\bibitem{zthree} A. Font \etal \Journal{\NPB}{331}{421}{1990};
                J.A. Casas \etal \Journal{\NPB}{317}{171}{1989};
                D. Bailin \etal \Journal{\NPB}{298}{75}{1988};
                A.E. Faraggi \etal \Journal{\NPB}{335}{347}{1990};
                S. Chaudhuri \etal \Journal{\NPB}{469}{357}{1996};
\bibitem{eu}A.E. Faraggi, \Journal{\PLB}{278}{131}{1992}.
\bibitem{slm}A.E. Faraggi, \Journal{\NPB}{387}{239}{1992}; hep-th/9707112.
\bibitem{FFF}K. Kawai \etal \Journal{\NPB}{288}{1}{1987};
             I. Antoniadis \etal \Journal{\NPB}{289}{87}{1987}.
\bibitem{ztwo}A.E. Faraggi \Journal{\PLB}{326}{62}{1994}.
\bibitem{efn}J. Ellis \etal \Journal{\PLB}{419}{123}{1998};
\bibitem{CKM}A.E. Faraggi and E. Halyo, \Journal{\NPB}{416}{63}{1994}.
\bibitem{CF}G.B. Cleaver and A.E. Faraggi, hep-ph/9711339.
\bibitem{FP2}A.E. Faraggi and J.C. Pati, hep-ph/9712516.
\bibitem{univ}A.E. Faraggi, hep-ph/9801409.
\bibitem{flatanalysis} J.L. Lopez and D.V. Nanopoulos, 
\Journal{\PLB}{251}{73}{1990};
A.E. Faraggi, \Journal{\NPB}{403}{101}{1993};
G.B. Cleaver \etal \Journal{\PRD}{77}{2701}{1998}.
\bibitem{dsp} Z. Lalak,  hep-th/9708410; N. Arkani-Hamed, M. Dine and S.P.
Martin, hep-ph/9705270; N. Irges, hep-ph/9805237.
\end{thebibliography}
\end{document}